\documentclass[aps,prb,twocolumn,superscriptaddress]{revtex4-1}
\usepackage{hyperref}
\usepackage{url}
\usepackage{graphicx}
\usepackage{bm}
\usepackage{amssymb}
\usepackage{amsmath}
\usepackage[version=3]{mhchem}
\usepackage[dvipsnames]{xcolor}

\newcommand\myshade{85}
\colorlet{mylinkcolor}{violet}
\colorlet{mycitecolor}{YellowOrange}
\colorlet{myurlcolor}{Aquamarine}

\hypersetup{
  linkcolor  = mylinkcolor!\myshade!black,
  citecolor  = mycitecolor!\myshade!black,
  urlcolor   = myurlcolor!\myshade!black,
  pdftitle={Physically founded phonon dispersions of few-layer materials, and the case of borophene},
  colorlinks = true
}


\begin{document}
\onecolumngrid
\begin{center}
  \includegraphics{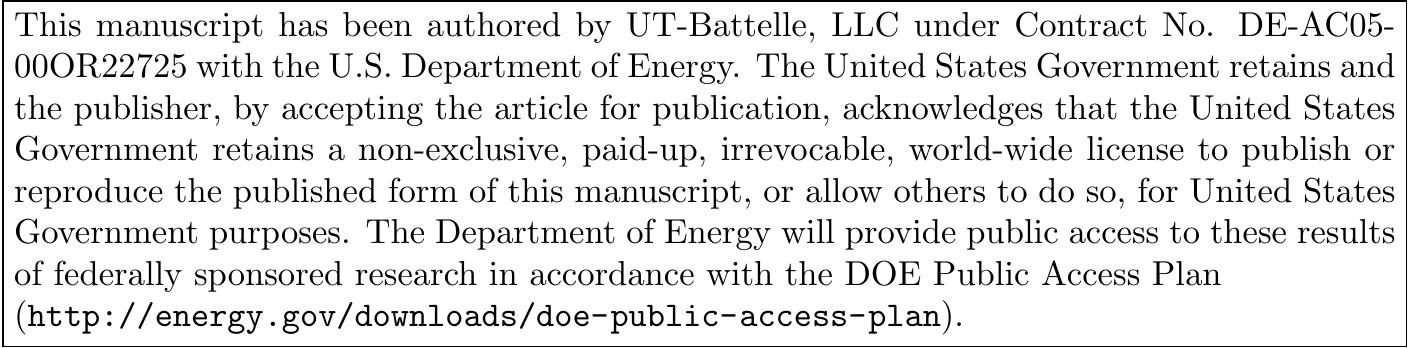}
\end{center}
\newpage

\title{Physically founded phonon dispersions of few-layer materials, and the case of borophene}
\author{Jes\'us Carrete}
\affiliation{CEA, LITEN, 17 Rue des Martyrs, 38054 Grenoble, France}
\author{Wu Li}
\affiliation{CEA, LITEN, 17 Rue des Martyrs, 38054 Grenoble, France}
\author{Lucas Lindsay}
\affiliation{Materials Science and Technology Division, Oak Ridge National Laboratory, Oak Ridge, Tennessee 37831, USA}
\author{David A. Broido}
\affiliation{Department of Physics, Boston College, Chestnut Hill, Massachusetts 02467, USA}
\author{Luis J. Gallego}
\affiliation{Departamento de F\'isica de la Materia Condensada, Facultad de F\'isica, Universidad de Santiago de Compostela, E-15782 Santiago de Compostela, Spain}
\author{Natalio Mingo}
\email{natalio.mingo@cea.fr}
\affiliation{CEA, LITEN, 17 Rue des Martyrs, 38054 Grenoble, France}
\begin{abstract}
An increasing number of theoretical calculations on few-layer materials have been reporting a non-zero sound velocity for all three acoustic phonon modes. In contrast with these reports, here we show that the lowest phonon dispersion branch of atomistically described few-layer materials should be quadratic, and this can have dramatic consequencies on calculated properties, such as the thermal conductivity. By reformulating the interatomic force constants (IFC) in terms of internal coordinates, we find that a delicate balance between the IFCs is responsible for this quadraticity. This balance is hard to obtain in \textit{ab-initio} calculations even if all the symmetries are numerically enforced \textit{a posteriori}, but it arises naturally in our approach. We demonstrate the phenomenon in the case of borophene, where a very subtle correction to the \textit{ab-initio} IFCs yields the physically correct quadratic dispersion, while leaving the rest of the spectrum virtually unmodified. Such quadraticity nevertheless has a major effect on the computed lattice thermal conductivity, which in the case of borophene changes by more than a factor 2, and reverses its anisotropy, when the subtle IFC correction is put in place.
\end{abstract}

\maketitle

\section{Introduction}
Exfoliated single-layer and few-layer materials have recently emerged as potentially revolutionary for many different applications \cite{geim_van_2013}. Following the first isolation of graphene, the number of exfoliated 2D materials to be studied continues to increase every year. Some recent examples drawing large audiences are  \cite{wang_electronics_2012} \ce{MoS2} and black phosphorene \cite{liu_phosphorene:_2014,shulenburger_nature_2015}, which are particularly promising for electronics and optoelectronics. Other few-layer materials, such as silicene, are still just at the theoretical modeling stage, but large experimental efforts are currently underway to bring them into real applications.\cite{svec_silicene_2014,vogt_silicene:_2012} An exciting development has just taken place with the first experimental synthesis of borophene,\cite{mannix_synthesis_2015} whose existence had been theoretically predicted not long before. \cite{tang_novel_2007,piazza_planar_2014,zhou_semimetallic_2014}

The 2D character of these systems brings in unique physical properties, sometimes resulting in desirable qualities, such as high electron mobilities, or high thermal conductivities  \cite{Butler_Progress_2013,miro_atlas_2014}. A central role in determining many of these properties is played by the phonon dispersion. Anharmonic phonon scattering, and electron-phonon coupling, fundamentally affect the thermal and electrical conductivities of the material, and their magnitudes are directly related to the features of the phonon dispersion, particularly the lower frequency spectrum. The existence or not of a quadratic acoustic phonon branch in the spectrum can completely change the physics of the problem at hand. It can for example lead to appreciable changes in the thermal conductivity, which has been an object of debate in the particular case of graphene \cite{bonini_acoustic_2012,lindsay_phonon_2014,fugallo_thermal_2014}.

It is therefore surprising that no consensus exists as to whether unstrained few-layer systems should always display one flexural phonon branch with quadratic phonon dispersion at long wavelengths, even though this is what elasticity theory predicts \cite{Landau_theory_1986}. Indeed, many recent \textit{ab-initio} calculations report three linear-dispersion acoustic branches (silicene \cite{manjanath_mechanical_2013,xie_thermal_2014, gu_first-principles_2015}, phosphorene \cite{jain_strongly_2015}, \ce{MoS2} \cite{fan_structural_2015}, \ce{WS2} \cite{berkdemir_identification_2013}, \ce{MgB6} \cite{xie_novel_2014}, \ce{WSe2} \cite{zhou_first-principles_2015}) whereas some other \textit{ab-initio} calculations, and virtually every empirical potential calculation, report one quadratic and two linear acoustic branches (silicene \cite{zhang_thermal_2014}, phosphorene \cite{qin_anisotropic_2015,zhu_coexistence_2014}, \ce{MoS2} \cite{molina-sanchez_phonons_2011}, \ce{WS2} \cite{molina-sanchez_phonons_2011}). Arguments have even been given to suggest that in a buckled system the flexural branch dispersion should no longer be quadratic \cite{gu_first-principles_2015}.

In order to clarify the situation, in this paper we formulate a necessary condition for the force constants of any relaxed mechanical system to be physically correct. This formulation automatically guarantees the fulfillment of the constraints due to crystal symmetry, translational invariance and rotational invariance \cite{born_dynamical_1998,gazis_conditions_1966,wang_new_2007}.  Using this approach we show that a quadratic dispersion acoustic branch is always present on suspended few-layer systems. We also show that this can have a strong impact in the calculated thermal conductivity of some 2D systems, and we illustrate it in the case of the Pmmn borophene structure, arguably one of the most complex 2D materials studied \textit{ab-initio} to date.

\section{Theory}
Our starting premise is: The total potential energy of a physical system of particles, in the absence of any external fields, can always be expressed as a function of a set of internal coordinates that are scalars. Such coordinates can include interatomic distances, angles, dihedrals, and so on. This principle is a consequence of the fundamental symmetries of physics. It is also intuitively evident, as the internal coordinates fully determine the positions of atoms relative to each other, while guaranteeing the isotropy and homogeneity of space. This principle is fully general and not related to any particular form of the energy itself.

If the system is close enough to a point of static mechanical equilibrium, its energy can be expanded around the equilibrium positions: $E=\sum_{ij}{\partial^2 E\over \partial x_i\partial x\j}\Delta x_i\Delta x_j$, where the $\{x_i\}$ are Cartesian coordinates. Since $E$ depends only on the internal coordinates, $l_n$, we can write:

\begin{equation}
{\partial^2 E\over \partial x_i\partial x_j} = \sum_{l,l'}{\partial l'\over \partial x_j}{\partial^2 E\over \partial l\partial l'}{\partial l\over \partial x_i}.
\label{eqn:change}
\end{equation}
Since the derivatives ${\partial l\over \partial x_i}$ are known, this provides us with the most general method to generate physically correct interatomic force constants. It suffices to first choose the equilibrium positions, from them compute all the internal coordinate derivatives, and then assign values to the energy derivatives with respect to all pairs of internal coordinates.
To guarantee that the system is dynamically stable, the ${\partial^2 E\over \partial l\partial l'}$ needs to be symmetric and positive definite. This can be ensured by generating it as the square of a matrix: ${\partial^2 E\over \partial l\partial l'}=\sum_k A^T_{lk}A_{kl'}$.

Such a representation is overcomplete, however: different choices of the ${\partial^2 E\over \partial l\partial l'}$ may yield the same force constants. This is easily resolved by the fact that all interatomic distances can be expressed as sole functions of a minimal set of $d\times n$ internal coordinates. For $n$ atoms in 3D, it suffices to choose three atoms and specify their three internal coordinates plus the $3(n-3)$ internal coordinates relating them to the remaining atoms, resulting in $3n-6$ values. Similarly in strict 2D we need to choose two reference atoms, and consider only $2n-3$ values. This has an important consequence: in 3D (2D) 6 (3) of the system's vibrational eigenvalues are always 0, and correspond to the rigid translations and rotations of the system.

Thus formulated, the approach is valid for atomic clusters or molecules. To apply it to periodic systems, we shall consider (as it is standard practice) that the system is a periodic array of supercells which interact only with those supercells adjacent to them. Thus, the energy can be fully expressed as a sum over adjacent pairs of supercells, $p$ (with $p$ also including the case of interactions within the same cell.) The force constants can then be written in the fashion previously described, as
\begin{equation}
H_{i,j}=\sum_p \sum_{l,l'}{\partial l'\over \partial x_j}W^{(p)}_{l,l'}{\partial l\over \partial x_i},
\label{eqn:periodic}
\end{equation}
where $W^{(p)}_{l,l'}={\partial^2 E\over \partial l\partial l'}$, for $l,l'\in\{p\}$, and $p$ denotes the pair of neighboring supercells.
Since the system is periodic, $W^{(p)}_{l,l'}$ must be defined in such a way that it does not change if the atoms involved in the definition of internal coordinates $l$ and $l'$ are simultaneously translated by a lattice vector.

The approach described allows for the random generation of arbitrary  atomic positions and force constants, which satisfy the relevant symmetries ``by construction''. A simple illustrative example is that of a 2D infinite chain with motions restricted to the (x,y) plane, and 2 atoms in the unit cell, and interactions to 3rd nearest neighbors. The relative position of the two atoms is random, accounting for 2 arbitary parameters. Another 5 random parameters correspond to the $W_{l,l'}$ between the atoms in two adjacent unit cells. We can then numerically compute the phonon dispersion near $\Gamma$ ($q = 0$), and evaluate the exponent $n$ of the frequency dependence on wave number $\omega\propto q^n$ for the lowest acoustic branch. Within 100000 random configurations, all of them had $n$ compatible with $2$, meaning that every system contains a quadratic dispersion.

The analysis would not be complete, however, unless we can address the phonon dispersion of real few-layer systems calculated from first principles, and ascertain whether they always contain a quadratic acoustic mode. The approach is to find the set of $W_{l,l'}$ that, using Eq. \eqref{eqn:periodic}, produce a set of physical IFCs that is closest to the raw (i.e. unphysical) \textit{ab-initio} calculated IFCs. As we will see, in every case studied just very small changes of some IFCs are able to turn the set into a physical one, resulting in dispersions that virtually do not differ from the raw ones, except for the perfect quadraticity of the lowest acoustic mode.

The procedure to find the closest physical set of IFCs for a given equilibrium system is as follows. First of all, the raw IFCs, $H^{\alpha\beta}_{0,j}$ are calculated, between the $\alpha$ and $\beta$ degrees of freedom of the central unit cell and its $j^{th}$ neighbouring unit cell, respectively. The standard enforcement of translational invariance ensures that the three acoustic phonon modes pass through zero at the $\Gamma$ point ($\vec q=(0,0,0)$.) However, in general, the three modes have a linear component, as shown in Fig. \ref{fig:1}.

Then we define a suitable set of internal coordinates for the extended system. We only need to consider the atoms in the central unit cell, and those interacting with them within neighboring cells. We thus pick 3 of those atoms, making sure they are not collinear, which will act as a base, and we take their three interatomic distances as coordinates. We could in principle continue picking one of the remaining atoms at a time, and selecting its 3 distances to each of the basis atoms as coordinates. However, this approach runs into problems if the new atom lies in the same plane as the triangle defined by the three basis atoms. In such case, the internal coordinates thus defined cannot describe out-of-plane motions. To avoid this problem, when the new atom is coplanar with the basis we choose a different set of coordinates corresponding to
\begin{subequations}
\begin{align}
y^{(j)}_1 = |\vec x_j-\vec x_1|,\\
y^{(j)}_2= \vec d_{j,1}\cdot (\vec d_{2,1}\times \vec d_{3,2}),\\
y^{(j)}_3= \vec d_{j,1}\cdot[(\vec d_{3,2}\times \vec d_{2,1})\times \vec d_{3,2}],
\end{align}
\end{subequations}
where $\vec d_{i,j}=\vec x_i-\vec x_j$. There are many possible choices other than the above one. From now on we will label internal coordinates with a single index, as $y_l$, rather than two.

The next step is to construct the matrix $\bf{\eta}$ relating the IFCs in Cartesian and internal coordinate representations. In other words, we want to express the force constants as
\begin{equation}
\bf{H_{i,j}}=\sum_{l,l'}\eta_{(i,j),(l,l')} \bf{W_{(l,l')}}.
\label{eq:minsq}
\end{equation}
Thus,
\begin{equation}
\eta_{(i,j),(l,l')} = (\nabla_{\vec x_i} y_l) \otimes (\nabla_{\vec x_j} y_{l'}).
\end{equation}
$\eta$ is non zero when atoms $i$ and $j$ are involved in the internal coordinates $y_l$ and $y_{l'}$. This means that $\eta$ is a sparse object, which permits us to take advantage of sparse matrix arithmetic in the calculation. However, it is important to realize that the value of $\bf {W}_{(l,l')}$
will affect not only each pair of atoms $(i,j)$ belonging to $y_l$ and $y_{l'}$, but also all those pairs equivalent to the first one by a symmetry operation of the space group (Bravais lattice translations and point group symmetries). Thus, in the process of constructing $\eta$, these equivalent pairs $\tilde i,\tilde j$ have to be identified, and the values of  $\eta_{(\tilde i,\tilde j),(l,l')}$ properly set alongside with those of $\eta_{(i,j),(l,l')}$.

The last step consists in solving Eq. \eqref{eq:minsq} by least squares. In practical terms, this was done by considering $H$ and $W$ as sparse arrays, and $\eta$ as a sparse matrix. The newly obtained $H$ is the closest one to the original set that satisfies the conditions required for the system to be physical.

\section{Application to borophene phonon dispersions}
The result of the approach just described is illustrated for a novel quasi-two-dimensional material: borophene. The possibility of a boron-based analogue of graphene -- borophene -- had recently been theoretically predicted \cite{tang_novel_2007,piazza_planar_2014,zhou_semimetallic_2014} and its first experimental demonstration has been just reported \cite{mannix_synthesis_2015}. Boron is carbon's neighbor in the periodic table and has similar valence orbitals, but its electron deficiency prevents it from forming graphene-like planar honeycomb structures. The study of boron and its compounds is singularly difficult, to the point of being considered by some authors as ``arguably the most complex element in the periodic table'' \cite{oganov_boron:_2009}. Many proposed new phases of boron have been revealed as being actually boron-rich compounds with very complex crystal structures \cite{witness_material_2010}. In fact, boron displays multiple bulk phases \cite{oganov_ionic_2009,witness_material_2010}, and the formation of clusters, fullerene-type cages and nanotubes made of B is also a subject of ongoing debate \cite{kiran_planar--tubular_2005,huang_concentric_2010,ciuparu_synthesis_2004,oger_boron_2007,gonzalez_szwacki_b80_2007,yang_ab_2008,singh_probing_2008,de_energy_2011}. The ambient-conditions stable phase of B was only recently determined through a series of computational studies, the earliest of which \cite{van_setten_thermodynamic_2007} only dates back to 2007. It comes as little surprise, then, that the study of the 2D allotropes of boron has also progressed through iterative improvement upon previous results.

\begin{figure}[htb]
  \begin{center}
    \noindent\includegraphics*[width=0.95\columnwidth]{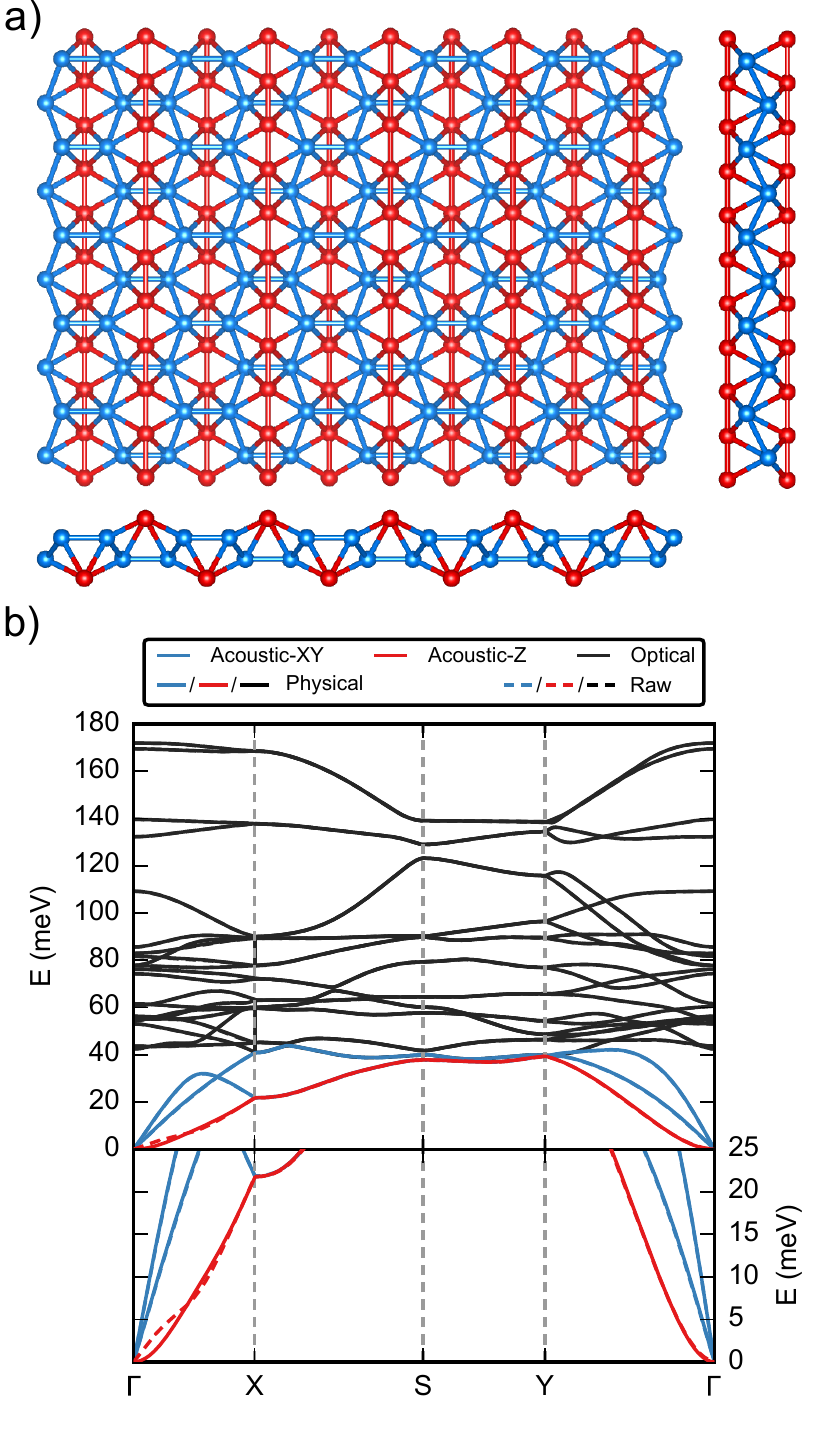}    
    \caption{(a) Atomic structure of Pmmn boron, showing 25 unit cells. All atoms belong to the same element; colors are only intended to guide the eye. (b) Comparison between raw (dashed) and physical (solid) phonon dispersions.}
    \label{fig:1}
  \end{center}
\end{figure}

Density functional theory (DFT) calculations \cite{tang_novel_2007} suggest that single-atomic-layer B sheets composed of triangular and hexagonal motifs are locally stable, the most stable structure of this kind being called the $\alpha$-sheet. Theoretical and experimental studies on the quasi-planar \ce{B_{36}} cluster with a central hexagonal hole seem to support this view \cite{piazza_planar_2014}. However, more recent calculations \cite{zhou_semimetallic_2014} have predicted two novel 2D boron structures with non-zero thickness that are considerably more stable than the $\alpha$-sheet. These two structures represent a more radical departure from the graphene prototype than silicene, germanene and their binary relatives. Nevertheless, one of these phases, belonging to space group Pmmn, also shows a distorted Dirac cone in its electronic band structure. Hence, this phase is extremely interesting as the first material with massless fermions that is not closely related to a graphene-like honeycomb structure \cite{zhou_semimetallic_2014}.

Pmmn borophene has a buckled structure, containing 8 atoms in the unit cell, and being considerably more complex than those of other 2D materials like graphene or phosphorene for example (see Fig. \ref{fig:1}). Therefore, it is particularly well suited to illustrate the existence of a quadratic phonon branch in finite thickness 2D materials. Details of the \textit{ab-initio} calculations are provided in the supplemental information. Our fully relaxed orthorhombic unit cell has side lengths $a=4.52\,\mathrm{\AA}$ and $b=3.25\,\mathrm{\AA}$, almost identical to Zhou \textit{et al.}'s result \cite{zhou_semimetallic_2014}. The difference between the maximum and minimum atomic positions along the $OZ$ axis is $2.21\,\mathrm{\AA}$. The slab is formed by four atomic layers along this axis, with an average distance between them of $0.74\,\mathrm{\AA}$.

\begin{figure}[htb]
  \begin{center}
    \noindent\includegraphics*[width=\columnwidth]{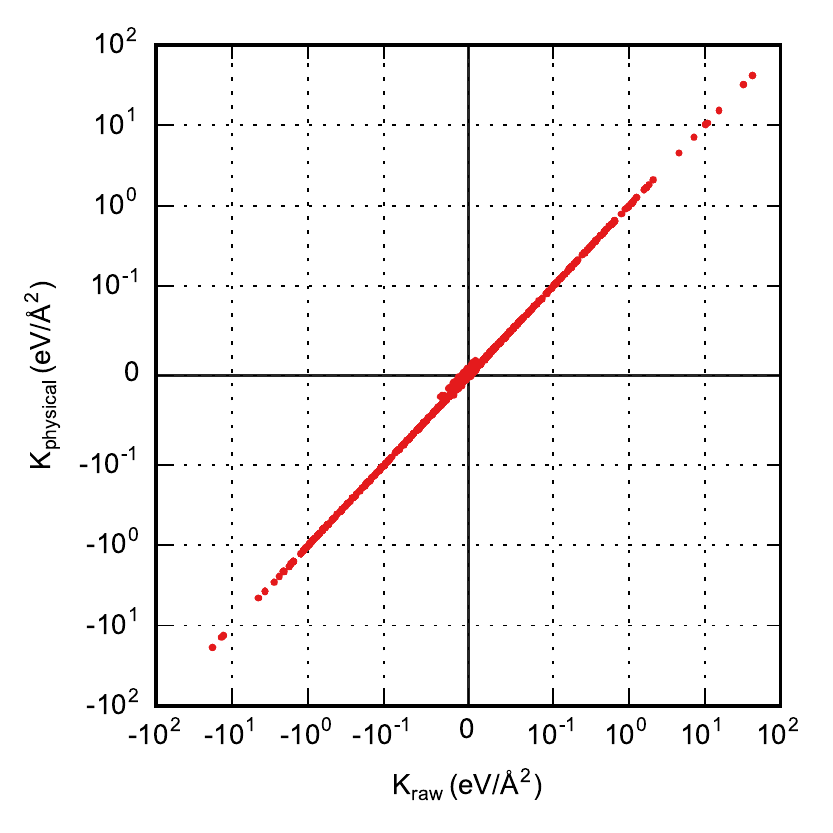}
    \caption{Comparison between raw (ordinate) and physical (abscissa) IFC values. (Symmetric logarithmic axes with linear threshold at $10^{-1}\,\mathrm{eV/\AA^2}$.)}
    \label{fig:2}
  \end{center}
\end{figure}

The raw calculated phonon dispersion displays a clear linear mode along the $\Gamma$-X direction. By ``raw" we mean calculated in the conventional fashion, where the IFCs that result from the \textit{ab-initio} calculation are only corrected by enforcement of translational invariance, as implemented \textit{e.g.} in Phonopy \cite{phonopy}. Application of the previously described approach results in a negligible change of the dispersion, except for turning the lowest acoustic mode into a perfectly quadratic dispersion near $\Gamma$ (Fig. \ref{fig:1}). Looking at Fig. \ref{fig:2}, the change in IFCs is very small. Despite this, the effect on the frequency dependence of the lowest branch is dramatic.

 The same effect was obtained for all the systems we have tried. Any linearity of the lowest branch goes away, whereas the rest of the dispersion remains virtually unchanged. The changes in IFCs involved are really subtle, which is the reason why it is very difficult to get the correct quadratic dependence of the lowest branch by simply imposing the various symmetries numerically. The ``by construction'' approach described, on the contrary, is able to capture this subtle effect with high precision, since the physical constraints on the IFCs are built-in.

It has been reported that at finite temperature, the anharmonic coupling in a single 2D layer renormalizes the low frequency flexural branch in the long wavelength limit to $\omega_{ZA}(q)\propto q^\alpha$, with values for $\alpha$ ranging from about\cite{los_scaling_2009} $1.2$ to \cite{Mariani_flexural_2008} $3/2$ having been proposed. We note that this renormalizaton has nothing to do with the finite group velocity often obtained by \textit{ab-initio} calculations of zero temperature force constants, which is definitely an artifact. Any calculation of thermal conductivity should employ physically correct IFCs, which yield a quadratic dispersion for the lowest branch in the harmonic approximation. This branch could then be renormalized for anharmonic effects if so desired.

It is also important to mention that thermal fluctuations in 2D systems lead to ripples, and destroy long range order \cite{trzesowski_isothermal_2014}. The magnitude of this rippling effect depends on the system, and even on the growth conditions. Nevertheless, for wavelengths longer than the typical corrugation distance, the speed of sound of one of the three acoustic modes will still tend to zero, as predicted by elasticity theory.

An alternative procedure to the ``by construction'' approach proposed is to symmetrize the IFCs by enforcing all the different invariance conditions afterwards. This second approach seems to us more complicated, and less suitable than the ``by construction'' method, for the following reasons. The invariance of the potential energy with respect to translations, rotations, and crystal symmetry operations imposes constraints on the interatomic force constants.  However, as pointed out previously \cite{sluiter_determination_1998}, in typical supercell or DFPT  force constant calculations, such constraints are in general not satisfied because of intrinsic errors in the computational algorithms (\textit{e.g.} from incomplete basis sets or imperfect self-consistency). Therefore, these invariance constraints must be explicitly imposed on the force constants after they have already been calculated.  In a fully relaxed bulk crystal, the harmonic force constants must satisfy the Huang invariance conditions  \cite{born_dynamical_1998}, which gives $15$ independent constraint equations taking account of the symmetry.  In addition to these, there are $9k$ equations each imposing translational and rotational invariance, where $k$ is the number of atoms in a primitive unit cell.  These $18k+15$ conditions must be imposed, on top of those associated with the specific crystal symmetries.  Various approaches to do this have been implemented ($\chi^2$ minimization, Lagrange multipliers, and so on and so forth), but the best way to do this remains ambiguous (\textit{e.g.} allowing only a maximum percent change in IFCs \textit{vs.} an absolute change).

\section{Effect of corrected dispersion on thermal conductivity}

\begin{figure}[hbt]
 \begin{center}\
   \noindent\includegraphics*[width=\columnwidth]{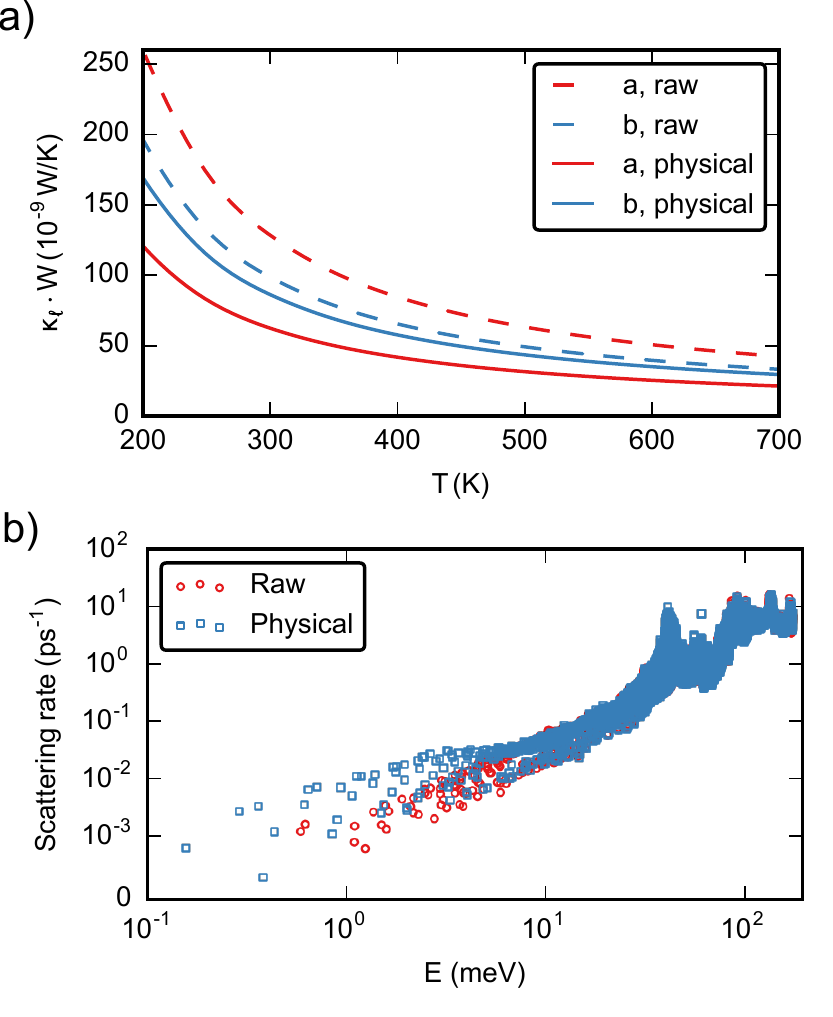}
   \caption{Comparison between calculated thermal conductivity (a) and scattering rates (b) using raw (red) or physical (blue) IFCs. The thermal conductivity of 2D systems has units of W/K. In order to compare with bulk materials one can divide the 2D conductivity by a chosen thickness of the film. For single or very thin layers, such thickness is ambiguous, however, so we have preferred to keep the 2D units here.}
 \label{fig:3}
 \end{center}
\end{figure}

The seemingly minor corrections of the phonon dispersion shown above can have a dramatic effect on the computed lattice thermal conductivity. In the case of borophene, the physically correct phonon dispersion yields $\kappa$ values which are less than half of those obtained with the unphysical linear dispersion from the uncorrected DFT calculation. Even more striking is the fact that the anisotropy is reversed: the correct result is $\kappa_b>\kappa_a$, whereas the linear dispersion calculation wrongly predicts $\kappa_a>\kappa_b$. This is related to the large enhancement of the low frequency phonon scattering rates, which is produced by the larger density of states of the quadratic branch. As shown in Fig. \ref{fig:3}, the phonon scattering rates are nearly unchanged, except for energies below $10\,\mathrm{meV}$, where the physical ones are roughly 10 times larger than the raw ones. It is these low frequency phonons that carry most of the heat, due to their longer lifetimes. Therefore, the effect on lattice thermal conductivity is substantial. We point out that the first experimentally sinthesized samples of borophene   show metallic characteristics,\cite{mannix_synthesis_2015} in which case an electronic contribution would add up to the total thermal conductivity. This metallic character might be linked to the fact that the samples are grown on a substrate, rather than suspended.

\section{Conclusions}
We have reformulated the expression for IFCs in terms of internal coordinates, thus ensuring that they are physically valid. In this formulation the lowest phonon branch of an atomistically described finite thickness layer is always quadratic in the harmonic approximation, in contrast with the linear dispersion sometimes found  in \textit{ab-initio} calculated phonon spectra of suspended layers. This quadraticity is linked to very subtle differences in the values of the IFCs, and it is hard to come by even if translational and rotational symmetry conditions are numerically enforced. On the contrary, in our formulation the IFCs are physically valid ``by construction'', and they automatically satisfy all the required symmetries without any need to impose them individually. By numerically finding the physical IFC set that is closest to the \textit{ab-initio} one, we obtained phonon dispersions that are almost indistinguishable from the original ones to the bare eye, except for the fact that their lowest phonon branch is quadratic at $\Gamma$. In the case of borophene this has a dramatic effect on the computed thermal conductivity, which decreases by 50\% and reverses its anisotropy when the correct physical force constants are used. We conclude that the linear dispersion of the out-of-plane modes often observed in \textit{\textit{ab-initio}} calculated spectra of 2D and few-layer materials is an artifact resulting from unphysical IFCs; any calculation dealing with phonons in 2D materials should ensure the physical correctness of the IFCs, as we have shown here.

\section*{Acknowledgments}
This work has been partly supported by the Air Force Office of Sponsored Research under grant No. FA9550-15-1-0187, the European Union's Horizon 2020 research and innovation programme under grant agreement No. 645776 (ALMA), and the M-Era programme through project ICETS.
L.J.G. acknowledes the support provided by the Spanish
Ministry of Science and Innovation (Grant No. FIS2012-33126) and by the Xunta de Galicia (AGRUP2015/11). L. L. acknowledges support from the U. S. Department of Energy, Office of Science, Office of Basic Energy Sciences, Materials Sciences and Engineering Division for work done at ORNL. D.A.B. acknowledges support from the NSF EFRI 2-DARE program under grant No. 1402949 and from ONR under grant No. N00014-13-1-0234.

\bibliographystyle{achemso}
\bibliography{bibliography}

\end{document}